# Increasing the Efficiency of Energy Scavengers * # +

S. M. Shahruz

Berkeley Engineering Research Institute
P. O. Box 9984
Berkeley, California 94709

**Abstract** In this paper, a methodology for designing efficient energy scavengers is proposed. The scavenger consists of a cantilever beam on which piezoelectric films and a mass are mounted. The mass at the tip of the beam is known as the proof mass and the device is called either an energy scavenger or a beam-mass system. The proof mass is a permanent magnet, where in its vicinity attracting permanent magnets are placed. It is shown that when the magnets have appropriate strengths and are placed appropriately, the vibration of the beam-mass system can be amplified, thereby the scavenged energy is increased. Examples are given throughout the paper.

**Keywords:** Energy scavengers, Beam-mass systems, Piezoelectric films, Permanent magnets, Nonlinear restoring spring force

## 1. Introduction

In the past two decades, researchers have been designing and studying devices by which energy from the environment can be scavenged. An example of energy scavenging is the conversion of the energy of vibration sources, which is usually neglected, into electricity to power micro-electronic devices; see, e.g., (Anton and Sodano, 2007), (Clark, 2005), (Roundy, et al., 2004), (Shahruz, 2006a), (Shahruz, 2006b), (Shahruz, 2006c), (Shahruz, 2007a), (Shahruz, 2007b), and references therein. Examples of vibration sources, the energy of which to be scavenged, are buildings, bridges, cars, trains, aircraft, ships, manufacturing tools, etc.

---

* There is a provisional patent application based on the work presented in this paper.

A device that scavenges energy efficiently from the environment is called an energy scavenger. There are three transduction principles based on which the energy of vibration sources is converted into electricity: electrostatics, electromagnetism, piezoelectricity.

One type of energy scavenger designed based on piezoelectricity is shown in Fig. 1. This device consists of a cantilever beam on which piezoelectric films and a mass are mounted. The mass at the tip of the cantilever beam is known as the proof mass and the device is called either an energy scavenger or a beam-mass system. When a scavenger is mounted on a vibration source, say a panel, the cantilever beam would vibrate. The vibration of the beam is converted into electricity by the piezoelectric films. The scavenged energy is proportional to the amplitude of vibration of the tip of the beam; see, e.g., (Jeon, et al., 2005, p. 20). Hence, a means to increase the scavenged energy is to make the beam vibrate persistently with large amplitudes.

The goal of this paper is to show that it is possible to amplify the vibration of a beam-mass system by using a magnetic proof mass and placing carefully chosen attracting magnets in its vicinity. The organization of the paper is as follows. In Section 2, a mathematical model describing the dynamics of a beam-mass system is presented. In Section 3, first the dynamics of a magnet in an attracting magnetic field is studied. Then, the success of a magnetic proof mass and a magnetic field in its vicinity in amplifying the vibration of a beam-mass system is illustrated. Furthermore, it is explained how the magnetic field should be chosen. In Section 4, several comments regarding the proposed energy scavenger are made.

## 2. A mathematical model of vibrating beam-mass systems

A schematic of a beam-mass system is shown in Fig. 2. The length, width, and thickness of the beam are denoted by $l$, $w$, and $h$, respectively. The mass density and the modulus of elasticity of the beam are denoted by $\rho$ and $E$, respectively. The proof mass at the tip of the beam is assumed to be a point mass of mass $M$. The vibration source on which the cantilever beam is mounted exerts the acceleration $\ddot{u}(\cdot)$. Due to this external input, the beam vibrates transversally. The transversal displacement of the beam at an $x \in [0, l]$ and a $t \geq 0$ is denoted by $y(x, t) \in \mathbb{R}$.

With this setup, a mathematical model describing the dynamics of a beam-mass system is derived in (Shahruz, 2006c). This model is described briefly in the following. The transversal displacement of the beam is written as

$$y(x, t) = \phi(x)\, q(t)\,, \qquad (1)$$

for all $x \in [0, l]$ and $t \geq 0$. In Eq. (1), the real- and scalar-valued function $x \mapsto \phi(x)$, known as the trial function, is chosen as

$$\phi(x) = a(\alpha)(\frac{x}{l})^2 - b(\alpha)(\frac{x}{l})^3 , \tag{2}$$

for all $x \in [0, l]$, with

$$\alpha := \frac{M}{\rho whl} , \tag{3a}$$

$$a(\alpha) = \frac{\sin\lambda(\alpha) + \sin\mathrm{h}\lambda(\alpha)}{\cos\lambda(\alpha) + \cos\mathrm{h}\lambda(\alpha)}\,\lambda^2(\alpha) , \qquad b(\alpha) = \frac{1}{3}\,\lambda^3(\alpha) , \tag{3b}$$

where the dependence of $\lambda$ on $\alpha$ is given in (Karnovsky and Lebed, 2004, p. 188, Table 6.7(a)).

The real- and scalar-valued function $t \mapsto q(t)$ in Eq. (1), known as the generalized coordinate, is the solution of following linear second-order ordinary differential equation:

$$m\,\ddot{q}(t) + c\,\dot{q}(t) + k\,q(t) = -\,f\,\ddot{u}(t) , \qquad q(0) = 0 , \qquad \dot{q}(0) = 0 , \tag{4}$$

for all $t \geq 0$, where $c$ is a positive real number known as the damping coefficient, and

$$m := a_1(\alpha)M + a_2(\alpha)\rho whl , \qquad k := \frac{a_3(\alpha)Ewh^3}{3l^3} , \qquad f := a_4(\alpha)M + a_5(\alpha)\rho whl , \tag{5}$$

with

$$a_1(\alpha) := [a(\alpha) - b(\alpha)]^2 , \qquad a_2(\alpha) := \frac{a^2(\alpha)}{5} - \frac{2a(\alpha)b(\alpha)}{6} + \frac{b^2(\alpha)}{7} , \tag{6a}$$

$$a_3(\alpha) := a^2(\alpha) - 3a(\alpha)b(\alpha) + 3b^2(\alpha) , \tag{6b}$$

$$a_4(\alpha) := a(\alpha) - b(\alpha) , \qquad a_5(\alpha) := \frac{a(\alpha)}{3} - \frac{b(\alpha)}{4} . \tag{6c}$$

Using the value of $\alpha$ and the corresponding $\lambda(\alpha)$ in (Karnovsky and Lebed, 2004, p. 188, Table 6.7(a)), $a(\alpha)$ and $b(\alpha)$ in Eq. (3b) can be computed for all $\alpha \geq 0$. Having these quantities computed, it can be verified numerically that $0 < a_j(\alpha) < \infty$ for all $\alpha \geq 0$ and $j = 1, 2, \ldots, 5$.

The displacement of the tip of the beam is denoted by $y_l$. Following the steps in (Shahruz, 2006c), it is concluded that $y_l$ satisfies the following differential equation:

$$m\,\ddot{y}_l(t) + c\,\dot{y}_l(t) + k\,y_l(t) = -\,a_4(\alpha) f\,\ddot{u}(t)\,, \qquad y(0) = 0\,, \qquad \dot{y}(0) = 0\,, \tag{7}$$

for all $t \geq 0$ .

System (7) is a simple linear representation of the dynamics of beam-mass systems. It is shown in (Shahruz, 2007a) that, as far as the energy scavenging is concerned, this system adequately represents the dynamics of beam-mass systems. To system (7) there corresponds the linear restoring spring force $ky$ . In the following section, it will be explained how to alter this force nonlinearly in order to make the beam-mass system vibrate persistently with large amplitudes, and hence, increase the efficiency of energy scavenging.

## 3. An energy scavenger with magnets

In this section, the design of a beam-mass system whose efficiency in converting the energy of vibration sources into electricity is increased by means of permanent magnets is proposed. This section consists of several subsections. In the following, for the sake of brevity, the word magnet is used in place of permanent magnet.

### 3.1. Dynamics of a magnet in a magnetic field

In Fig. 3, two attracting magnets are shown. These magnets are separated from each other by a glass plate of thickness $\gamma$ . One magnet is fixed at the distant $d$ from the line $OX$ . The other one is free to move, where its distance from the line $OX$ is denoted by $y$ . The mass of the free magnet is $m_f$ .

By the theory of magnets (see, e.g., (Burke, 1986, p. 13), (Parker, 1990, p. 28)), the attractive force between the free and fixed magnets in the $y$ direction is

$$F_a(y) = \frac{\mu_f \mu\,(d - y)}{[(d - y)^2 + \gamma^2]^{3/2}}\,, \tag{8}$$

where $\mu_f > 0$ and $\mu > 0$ are the normalized pole strengths of the free and fixed magnets, respectively. Knowing $F_a(\cdot)$ , the motion of the free magnet in the $y$ direction is described by

$$m_f\,\ddot{y}(t) + c_g\,\dot{y}(t) - \frac{\mu_f \mu\,(d - y(t))}{[(d - y(t))^2 + \gamma^2]^{3/2}} = 0\,, \qquad y(0) = y_0\,, \qquad \dot{y}(0) = y_1\,, \tag{9}$$

for all $t \geq 0$ , where $c_g > 0$ is the friction coefficient between the free magnet and the glass surface, and $y_0$ and $y_1$ are the initial displacement and velocity of the free magnet, respectively.

The equilibrium position of the free magnet is at $y_e = d$. This point is stable as it is shown in the following. Let the (candidate) Lyapunov function for system (9) be

$$V(t) = \frac{1}{2} m_f \, \dot{y}^2(t) + \frac{\mu_f \mu}{2\gamma} - \frac{\mu_f \mu}{2\,[(d - y(t))^2 + \gamma^2]^{1/2}} \,, \tag{10}$$

for all $t \geq 0$. It is clear that $V(\cdot)$ is a positive definite function.

The derivative of $V(\cdot)$ with respect to time along the solution of system (9) is

$$\dot{V}(t) = -\frac{1}{2} c_g \, \dot{y}^2(t) \leq 0 \,, \tag{11}$$

for all $t \geq 0$. The function $\dot{V}(\cdot)$ is only negative semi-definite. Thus, to establish stability, LaSalle invariance principle (see, e.g., (Khalil, 2002, p. 128), (Sastry, 1999, p. 199)) is to be used. By this principle, it follows that the equilibrium position of the free magnet is globally asymptotically stable. That is, the free magnet moves towards the fixed magnet and eventually settles above it, where the distance of the two magnets is the shortest and is equal to $\gamma$. The stability of the equilibrium position of the free magnet that was just proved is, of course, obvious from the physical point of view.

With the insight gained regarding the motion of the free magnet caused by the attractive force of the fixed magnet, a high efficiency energy scavenger is designed.

### 3.2. A beam-mass system and magnets

In this subsection, the design of a novel energy scavenger with higher efficiency is presented. A schematic of this device is shown in Fig. 4(a). The scavenger is a cantilever beam with a proof mass at its tip. The proof mass, however, is now a magnet. The neutral (also known as the main) axis of the beam when unbent is denoted by $x$. It can be assumed without much error that when the beam bends, the trajectory of the proof mass is a circular arc of radius $l$. The displacement of the tip of the beam is denoted by $y$. Suppose that dimensions of the beam, $l$, $w$, and $h$, and mass of the proof mass, $M$, are designed such that the beam-mass system has a desired resonant (fundamental) frequency. Methods in (Shahruz, 2006c) can be used to achieve such a design.

In the vicinity of the proof mass, $2n$ magnets of different strengths are placed. The placement of the magnets is symmetric with respect to the $x$-axis. More precisely, the $n$ magnets above and the other $n$ below the $x$-axis are the mirror image of each other geometrically and with respect to the strengths of the magnets. The $i$-th fixed magnet for an $i = 1, 2, \ldots, n$ and its mirror image magnet lie on a circular arc of radius $l + \gamma_i$, where $\gamma_i$ is a positive real number. The vertical distance of the $i$-th magnet from the $x$-axis and that of its mirror image magnet is denoted by $d_i$.

Using the force in Eq. (8) and geometric relations in Fig. 4(b), the attractive force applied by the fixed magnets to the magnetic proof mass is

$$F_a(y) = sgn(y) \sum_{i=1}^{n} \frac{\mu_f \mu_i (d_i - |y|)}{[(d_i - |y|)^2 + (((l+\gamma_i)^2 - d_i^2)^{1/2} - (l^2 - y^2)^{1/2})^2]^{3/2}}$$

$$- sgn(y) \sum_{i=1}^{n} \frac{\mu_f \mu_i (d_i + |y|)}{[(d_i + |y|)^2 + (((l+\gamma_i)^2 - d_i^2)^{1/2} - (l^2 - y^2)^{1/2})^2]^{3/2}}, \qquad (12)$$

for all $y \in \mathbb{R}$, where $\mu_f > 0$ is the normalized magnetic strength of the proof mass and $\mu_i > 0$ for an $i = 1, 2, \ldots, n$ is the normalized magnetic strength of the $i$-th fixed magnet and that of its mirror image magnet. In Eq. (12), $sgn(\cdot)$ is the signum function given by $sgn(y) = 1$ for $y > 0$, $sgn(y) = -1$ for $y < 0$, and $sgn(0) = 0$. It is clear that $y \mapsto F_a(y)$ is an odd function. The first (respectively, second) term in Eq. (12) represents the summation of the attractive forces of the magnets placed above (below) the $x$-axis when the proof mass is above this axis.

Knowing $F_a(\cdot)$, the displacement of the tip of the beam is now governed by the following nonlinear second-order differential equation:

$$m\,\ddot{y}(t) + c\,\dot{y}(t) + F_s(y(t)) = -\,a_4(\alpha) f\,\ddot{u}(t), \qquad y(0) = 0, \qquad \dot{y}(0) = 0, \qquad (13)$$

for all $t \geq 0$, where $c$ is the damping coefficient of the beam, $m$ and $f$ are the same as those in Eq. (4), $a_4(\alpha)$ is that in Eq. (6c), and the function $F_s : \mathbb{R} \to \mathbb{R}$ is given by

$$F_s(y) = ky - sgn(y) \sum_{i=1}^{n} \frac{\mu_f \mu_i (d_i - |y|)}{[(d_i - |y|)^2 + (((l+\gamma_i)^2 - d_i^2)^{1/2} - (l^2 - y^2)^{1/2})^2]^{3/2}}$$

$$+ sgn(y) \sum_{i=1}^{n} \frac{\mu_f \mu_i (d_i + |y|)}{[(d_i + |y|)^2 + (((l+\gamma_i)^2 - d_i^2)^{1/2} - (l^2 - y^2)^{1/2})^2]^{3/2}}, \qquad (14)$$

with $k$ the same as that in Eq. (4). The function $F_s(\cdot)$ represents the restoring spring force of the beam. Due to the magnetic proof mass and the magnets in its vicinity, the spring force of the beam is now effectively nonlinear.

The function $F_s(\cdot)$ plays a crucial role in the design of efficient energy scavengers. In the following, it will be shown that: (i) for certain spring force functions $F_d(\cdot)$ in place of $F_s(\cdot)$, system (13) exhibits persistent vibration with large amplitudes; (ii) by appropriate choices of $\mu_i$, $d_i$, and $\gamma_i$ for all $i = 1, 2, \ldots, n$ in Eq. (14), it is possible to approximate a function $F_d(\cdot)$ by $F_s(\cdot)$ with reasonable accuracy; that is, to realize a desired nonlinear spring force by a collection of magnets.

To illustrate that an appropriate spring force function $F_d(\cdot)$ can make a beam-mass system vibrate persistently with large amplitudes, an example is given.

**Example 3.1:** In systems (7) and (13), let

$$m = 1\,, \qquad c = 0.1\,, \qquad k = 1\,, \qquad a_4(\alpha) f = 1\,. \tag{15}$$

Furthermore, let the function $F_d : \mathbb{R} \to \mathbb{R}$ given by

$$F_d(y) = \frac{y\,[(4 - y^2)^2 + 0.1\, y^2]}{[(4 - y^2)^2 + 16\, y^2]}\,, \tag{16}$$

replace $F_s(\cdot)$ in Eq. (13). The graph of the nonlinear spring force $y \mapsto F_d(y)$ is shown in Fig. 5. Also, in this figure, the graph of the linear spring force $y \mapsto ky$ in system (7) with $k = 1$ is depicted. The external acceleration $\ddot{u}(\cdot)$ applied to systems (7) and (13) is considered to be a white noise of certain power whose amplitudes peak close to $\pm 2$.

With this setup, the linear beam-mass system (7) and the nonlinear beam-mass system (13) with $F_d(\cdot)$ in Eq. (16) in place of $F_s(\cdot)$ were simulated by using Simulink in Matlab (Matlab, 2007). Responses of systems (7) and (13) are shown in Figs. 6(a) and 6(b), respectively. It is evident that the amplitude of vibration of the tip of the nonlinear beam, $y$, is by far larger than that of the linear beam, $y_l$. It is also evident that the nonlinear beam-mass vibrates persistently.

To compare the intensity of vibration of the linear and nonlinear beam-mass systems, the following quantity is computed:

$$R_d := \frac{\|y\|_2}{\|y_l\|_2} = 2.59\,. \tag{17}$$

The power spectral density of $y_l(\cdot)$ and $y(\cdot)$ are shown in Fig. 7. From this figure, two conclusions are drawn: (i) the peak power of the response of the nonlinear beam-mass system is much larger than that of the linear system; (ii) the peak-power frequency of the response of the nonlinear beam-mass system is lower than that of the linear system; that is, the nonlinear beam is effectively softer (less stiff). □

**Remark:** For the purpose of energy scavenging, a beam-mass system must vibrate persistently with large amplitudes. More precisely, infrequent vibration of the tip of a beam, no matter how large, is not sufficient for energy scavenging. For instance, if the tip of a beam vibrates once or twice and soon settles at a large displacement and remains nearly still, then the conversion of the energy of vibration into electricity stops. □

### 3.3. Magnets that provide a desired nonlinear spring force

By Example 3.1, it is clear that a beam-mass system with the nonlinear spring force $F_d(\cdot)$ in Eq. (16) vibrates persistently with large amplitudes. The graph of this function is shown in Fig. 5. Suppose that a straight ray connecting a point on this graph to the origin is drawn. The slope of such a ray is called the instantaneous gain of the function $F_d(\cdot)$ at that point, or the instantaneous stiffness of the spring.

Consider now the points on the graph of $F_d(\cdot)$ whose abscissa are in the intervals $(-2, -0.75)$ and $(0.75, 2)$. Clearly, at these points the instantaneous gains of $F_d(\cdot)$ are less than $1$. Therefore, in these intervals, the beam behaves as a spring softer than the linear spring with the spring constant $k = 1$. A softer spring in certain intervals is the key to amplifying the vibration of the tip of a beam.

The function $F_d(\cdot)$, however, is only a mathematical function with no apparent relation to the magnets to be placed in the vicinity of the beam-mass system. The relation will be unraveled once the following question is answered: Can a collection of magnets (approximately) provide a desired nonlinear spring force $F_d(\cdot)$, such as that in Eq. (16)? Intuitively, the answer is yes. Suppose that due to an applied acceleration, the magnetic proof mass in Fig. 4(a) is moving upwards. Due to the attractive force of the fixed magnets in its vicinity, the proof mass will move faster and to a farther distance. This behavior can be interpreted as having a softer beam.

In this subsection, it is shown that by appropriate choices of $\mu_i$, $d_i$, and $\gamma_i$ for all $i = 1, 2, \ldots, n$ and for a reasonable $n$, the function $F_s(\cdot)$ in Eq. (14) can provide a good approximate of a desired nonlinear spring force $F_d(\cdot)$, such as that in Eq. (16). That is, by a collection of magnets that have appropriate strengths and are placed appropriately, a desired nonlinear spring force can be constructed. This construction is realized by solving the following minimization problem.

**Problem 3.1:** Consider a desired nonlinear spring force $F_d(\cdot)$ and the function $F_s(\cdot)$ in Eq. (14). The difference between these two functions is

$$e(y) = F_d(y) - F_s(y)\,, \tag{18}$$

for all $y \in \mathbb{R}$. Moreover, consider an interval $[-Y^*, Y^*]$ over which it is desired to have $F_d(\cdot)$ and $F_s(\cdot)$ close to each other.

Choose $n$ and determine $\mu_i$, $d_i$, and $\gamma_i$ in Eq. (14) for all $i = 1, 2, \ldots, n$, such that

$$\|e\|_{1,W} := \int_0^{Y^*} W(y)\,|\,e(y)\,|\,dy\,, \tag{19}$$

where $y \mapsto W(y)$ is a scalar-valued non-negative weighting function, is minimized. □

By solving Problem 3.1, the function $F_s(\cdot)$ can be constructed to be a reasonable approximate of $F_d(\cdot)$ over the interval $[0, Y^*]$, and by symmetry over $[-Y^*, Y^*]$. The weighting function $W(\cdot)$ can be chosen appropriately in order to penalize the difference between $F_d(\cdot)$ and $F_s(\cdot)$ in designated subintervals of $[0, Y^*]$. Constraints can be added to Problem 3.1 if design and fabrication considerations are to be taken into account. In solving Problem 3.1, some of the parameters $\mu_i$, $d_i$, and $\gamma_i$ may be chosen for some or all $i = 1, 2, \ldots, n$, while the remaining parameters are obtained by minimizing $\|e\|_{1,W}$. By using different norms, cost functions different from that in Eq. (19) can be defined.

Example are given to illustrate the possibility of constructing $F_d(\cdot)$ by a collection of magnets.

**Example 3.2:** Consider the desired nonlinear spring force $F_d(\cdot)$ in Eq. (16). In Eq. (14), let

$$n = 12\,, \quad l = 50\,, \quad k = 1\,, \quad \mu_f = 1\,, \quad \gamma_i = 0.35\,, \tag{20}$$

for all $i = 1, 2, \ldots, 12$. That is, all magnets are placed on a same circular arc of radius $50.35$.

The goal is to determine the strengths of each magnet, $\mu_i$, and its position form the $x$-axis, $d_i$, for all $i = 1, 2, \ldots, 12$, such that $F_s(\cdot)$ in Eq. (14) would be a good approximate of $F_d(\cdot)$. This goal is achieved by solving Problem 3.1 for the interval $[0, 4]$, while the weighting function $W \equiv 1$ over this interval. This problem was solved by using *fminsearch* in Matlab. The results are

$$\begin{aligned}
\mu_1 &= 0.0395\,, & d_1 &= 1.5962\,,\\
\mu_2 &= 0.0735\,, & d_2 &= 1.8702\,,\\
\mu_3 &= 0.1081\,, & d_3 &= 2.1192\,,\\
\mu_4 &= 0.1419\,, & d_4 &= 2.3552\,,\\
\mu_5 &= 0.1745\,, & d_5 &= 2.5817\,,\\
\mu_6 &= 0.2076\,, & d_6 &= 2.8020\,,\\
\mu_7 &= 0.2429\,, & d_7 &= 3.0194\,,\\
\mu_8 &= 0.2803\,, & d_8 &= 3.2360\,,\\
\mu_9 &= 0.3192\,, & d_9 &= 3.4524\,,\\
\mu_{10} &= 0.3663\,, & d_{10} &= 3.6693\,,\\
\mu_{11} &= 0.4778\,, & d_{11} &= 3.8985\,,\\
\mu_{12} &= 1.6750\,, & d_{12} &= 4.2622\,.
\end{aligned} \tag{21}$$

Using the values in Eqs. (20) and (21) in Eq. (14), the graph of $F_s(\cdot)$ is plotted in Fig. 8. For the purpose of comparison, the graph of $F_d(\cdot)$ in Eq. (16) is also plotted in this figure. It is clear that the graphs are reasonably close to each other. That is, by 24 magnets, it is possible to construct a reasonably accurate approximate of the desired nonlinear spring force $F_d(\cdot)$. It may be possible to make $F_s(\cdot)$ a better approximate of $F_d(\cdot)$ by choosing $n > 12$. However, then more parameters should be determined optimally at higher computational cost.

In Fig. 9, it is shown where the 24 magnets should be placed and how much their strengths should be. □

Next, the response of the beam-mass system in the magnetic field generated by the 24 magnets in Example 3.2 is determined.

**Example 3.3:** Let the parameters in system (13) be those in Eq. (15), and let the applied external acceleration $\ddot{u}(\cdot)$ be the same as that in Example 3.1. Furthermore, let $F_s(\cdot)$ in system (13) be that determined in Example 3.2.

With this setup, system (13) was simulated by using Simulink in Matlab. Response of this system, $y$, is shown in Fig. 10(a). The power spectral density of $y$ is shown in Fig. 10(b). It is evident that the amplitude of vibration of the tip of the beam in the magnetic field generated by the 24 magnets is larger than that of the linear beam.

The figure of merit $R_d$ for this example is

$$R_d = 2.84\,, \tag{22}$$

which even larger than that in Eq. (17). This figure implies that due to the magnets the beam-mass system vibrates with large amplitudes. □

The next example shows that a superb performance can be achieved with only a few magnets.

**Example 3.4:** Let the desired nonlinear spring force $F_d : \mathbb{R} \to \mathbb{R}$ be given by

$$F_d(y) = y[1 - \exp(-0.015y)]\,. \tag{23}$$

The graph of this function and that of the linear spring force $y \mapsto ky$ with $k = 1$ are shown in Fig. 11. Furthermore, in Eq. (14), let

$$n = 2\,, \quad l = 50\,, \quad k = 1\,, \quad \mu_f = 20\,. \tag{24}$$

The goal is to determine $\mu_i$ , $d_i$ , and $\gamma_i$ for $i = 1, 2$ , such that $F_s(\cdot)$ in Eq. (14) would be a good approximate of $F_d(\cdot)$ in Eq. (23). This goal is achieved by solving Problem 3.1 for the interval $[0, 7]$ , while the weighting function $W \equiv 1$ over this interval. This problem is solved by using *fminsearch* in Matlab. The results are

$$\mu_1 = 40\,, \quad d_1 = 7.14\,, \quad \gamma_1 = 11.32\,,$$

$$\mu_2 = 65\,, \quad d_2 = 15.02\,, \quad \gamma_2 = 7.43\,. \tag{25}$$

The graph of $F_s(\cdot)$ for the values in Eqs. (24) and (25) is plotted in Fig. 11. It is evident that $F_s(\cdot)$ and $F_d(\cdot)$ overlap. That is, by only four magnets, it is possible to construct a very accurate approximate of the desired nonlinear spring force $F_d(\cdot)$ . □

The response of the beam-mass system in the magnetic field generated by the four magnets in Example 3.4 is now determined.

**Example 3.5:** Let the conditions in Example 3.3 hold, except that $F_s(\cdot)$ in system (13) is that determined in Example 3.4. Response of system (13), $y$ , for this setup is shown in Fig. 12(a). The power spectral density of $y$ is shown in Fig. 12(b). Clearly, the amplitude of vibration of the tip of the beam in the magnetic field generated by the four magnets is much larger than that of the linear beam.

The figure of merit $R_d$ for this example is

$$R_d = 4.56\,, \tag{26}$$

which is much large than that in Eq. (17). □

### 3.4. A simple design

In the previous subsection, a methodology for designing efficient energy scavengers by means of magnets was given. It may be argued that: (i) determining the strengths and positions of the magnets by solving Problem 3.1 is not an easy task; (ii) fabricating a device with a collection of magnets of certain strengths that should be placed precisely in the vicinity of a beam-mass system is difficult. To simplify the design and fabrication, it is proposed to use only two magnets. It will be shown via an example that it is possible to bypass Problem 3.1, use only two magnets that are placed in the vicinity of a beam-mass system without stringent tolerances, but yet make the system vibrate persistently with large amplitudes.

**Example 3.6:** Let the parameters in system (13) be those in Eq. (15), and let the applied external acceleration $\ddot{u}(\cdot)$ be the same as that in Example 3.1.

Suppose that there are only two magnets placed in the vicinity of the magnetic proof mass; that is, $n = 1$ in Eq. (14). In this equation, let

$$\mu_f = 1\,, \qquad \mu_1 = 5.9\,, \qquad d_1 = 3\,, \qquad \gamma_1 = 1\,. \tag{27}$$

The graph of $y \mapsto F_s(y)$ for the values in Eqs. (27) is plotted in Fig. 13. Also, in this figure, the graph of the linear spring force $y \mapsto ky$ with $k = 1$ is depicted.

With this setup, system (13) was simulated by using Simulink in Matlab. Response of this system, $y$, is shown in Fig. 14(a). The power spectral density of $y$ is shown in Fig. 14(b). It is evident that the amplitude of vibration of the tip of the beam in the magnetic field generated by only two magnets is larger than that of the linear beam. In Fig. 13, for the points on the graph of $F_s(\cdot)$ whose abscissa are in the interval $(-3, 3)$, the instantaneous gains of $F_s(\cdot)$ are less than $1$. However, outside of this interval, the instantaneous gains are larger than $1$. This implies that the beam behaves as a stiff spring when the absolute value of the displacement of its tip is larger than $3$. Due to such stiffening, the amplitude of vibration rarely exceed $3$.

The figure of merit $R_d$ for this example is

$$R_d = 2.33\,. \tag{28}$$

This figure is not very large, however, it implies that due to only two magnets the beam-mass system vibrates with large amplitudes. □

The conclusion to be drawn from this section is that magnets in the vicinity of a magnetic proof mass can make beam-mass systems vibrate persistently with large amplitudes, thereby the scavenged energy from vibration sources would be larger.

## 4. Remarks regarding the proposed energy scavenger

In this section, several remarks regarding the proposed energy scavenger are made without details.

**1)** The desired nonlinear spring force $F_d(\cdot)$ plays an important role in increasing the energy scavenged by a beam-mass system. In Examples 3.1 and 3.5, it was shown that the functions $F_d(\cdot)$ in Eqs. (16) and (23), respectively, succeeded to amplify the vibration of a beam-mass system. It may seem that this function was chosen magically. Therefore, a question to be raised is: How should $F_d(\cdot)$ be chosen? A quick answer is: (i) the function $y \mapsto F_d(y)$ should be zero only at $y = 0$ ; (ii) the instantaneous gains of $F_d(\cdot)$ at some of its points are small.

An $F_d(\cdot)$ that does not satisfy (i) can lead to a collection of magnets some of which are so strong that would attract the magnetic proof mass and would not let it move. Hence, the beam-mass system would vibrate intensely, but infrequently. Energy scavenged from such a beam-mass system would not be much.

**2)** System (13) with spring force $F_s(\cdot)$ was simulated for many different random and deterministic inputs. In all cases, the vibration of the nonlinear beam-mass system was more intense than that of the corresponding linear beam-mass system, except when the input to the system was $\sin t$ . The reason is that the linear system (7) whose parameters are given in Eq. (15) has the resonant frequency $1\ rad/sec$ . Therefore, the input $\sin t$ makes the linear system resonate. It is thus inferred that the energy scavenger with magnets proposed in this paper is suitable for random vibration sources. An example of such a source is the surface of a road that applies a random force to tires of cars. For periodic vibration sources whose peak-power frequencies are low, the proposed energy scavenger can have a superb performance. For instance, in Example 3.5, the output of system (13) driven by $\sin 0.2t$ would have a large amplitude.

**3)** Several researchers have studied the vibration of a beam whose tip is close to two magnets; see, e.g., (Moon, 1987, p. 97) and references therein. The focus of such studies is on the chaotic behavior of the beam when it is excited by deterministic inputs. In their studies, the researchers were not concerned with the amplification of the beam vibration.

**4)** In (Shahruz, 2006a), (Shahruz, 2006b), and (Shahruz, 2006c), it is argued that in order to scavenge energy efficiently from a variety of vibration sources, an energy scavenger should have sufficient gain in designated frequency intervals. To achieve this goal, an ensemble of carefully designed beam-mass systems was proposed that would behave as a mechanical band-pass filter. Magnets can be added to such an ensemble of beam-mass systems for even better efficiencies.

**5)** Since the proposed energy scavenger uses magnets in the vicinity of a beam-mass system, one can venture to use coils of wire in order to induce current in them by the moving magnetic proof mass. That is, to jointly use the bending of the piezoelectric films and the motion of the moving magnet to convert the energy of a vibration source into electricity.

## 5. Conclusions

In this paper, a methodology for designing efficient energy scavengers was proposed. The scavenger consists of a cantilever beam on which piezoelectric films and a mass are mounted. The mass at the tip of the beam is known as the proof mass and the device is called either an energy scavenger or a beam-mass system. The energy scavenged by such a device is proportional to the amplitude of the beam vibration. Hence, a means to increase the scavenged energy is to make the beam vibrate persistently with large amplitudes. To achieve this goal, it was proposed to use a magnetic proof mass, where in its vicinity attracting magnets were placed. Introduction of magnets and devising a procedure for choosing them appropriately are the key contributions of the paper.

It was shown that when the magnets had appropriate strengths and were placed appropriately, the amplitude of vibration of the beam-mass system would be amplified by a factor of up to seven. Roughly speaking, due to the magnetic proof mass and the magnets in its vicinity, the spring force of the beam is effectively nonlinear, by which the beam behaves as a soft spring. Such a behavior is the reason for larger amplitudes of vibration. An effectively softer beam has another useful property: its fundamental frequency is low. This property is preferable due to the fact that the peak-power frequencies of vibration sources are usually low. A procedure for determining the strengths of the magnets and their positions to achieve desired nonlinear spring forces for beams was devised based on an minimization problem.

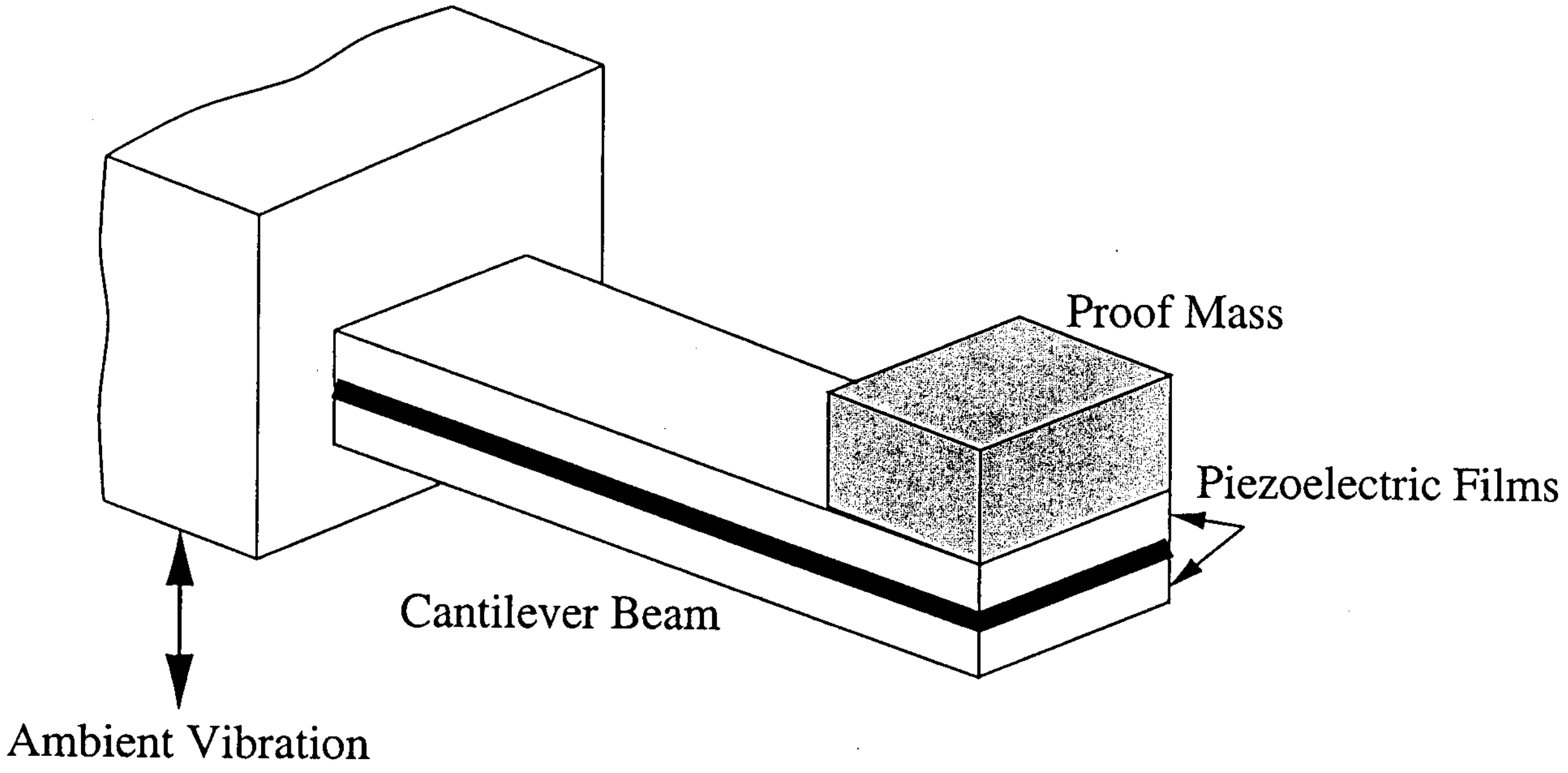

**Fig. 1:** A typical energy scavenger consists of a cantilever beam on which piezoelectric films and a mass, known as the proof mass, are mounted.

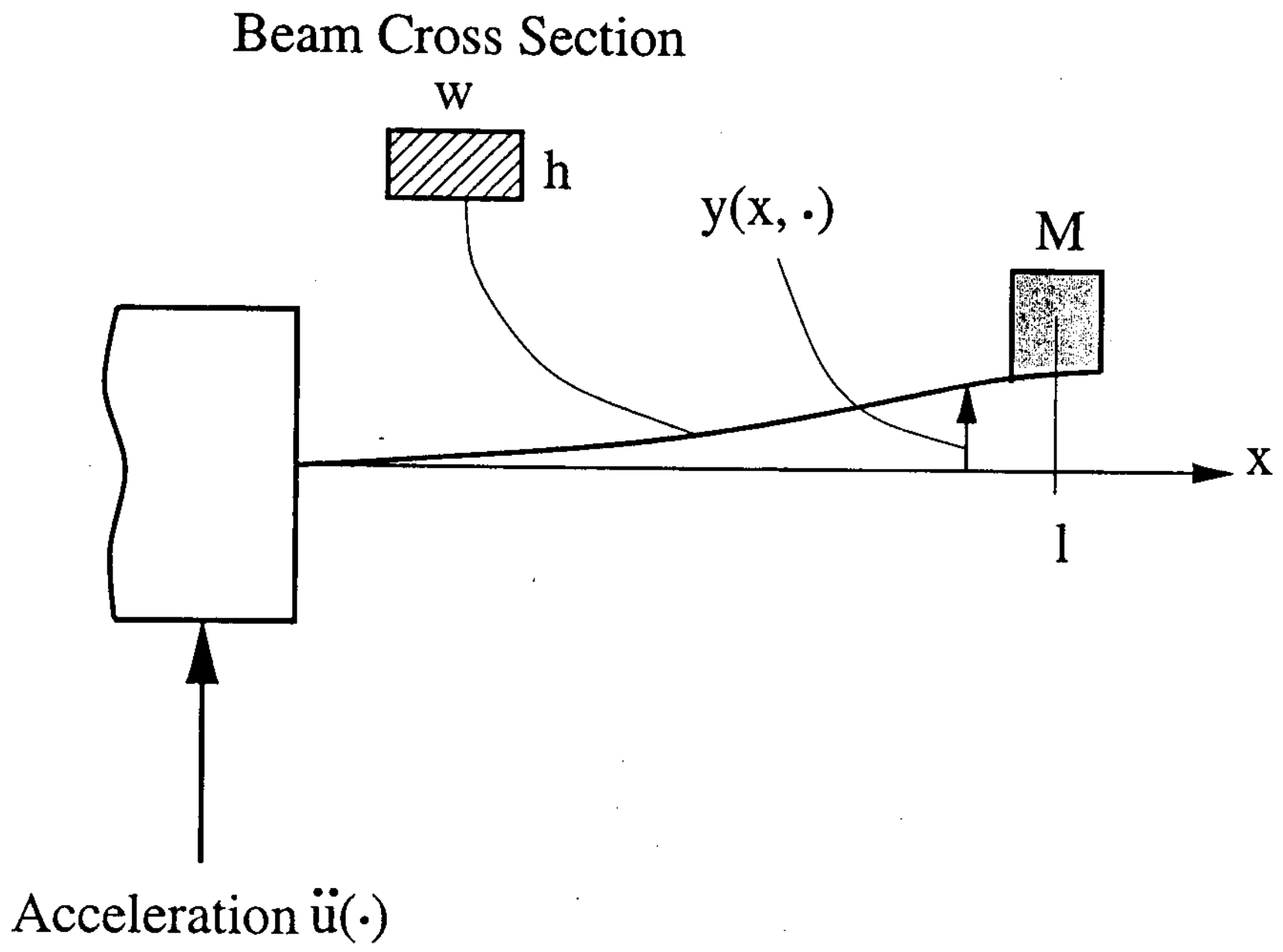

**Fig. 2:** A schematic of a beam with a proof mass at its tip. The vibration source exerts the acceleration $\ddot{u}(\cdot)$ . The transversal displacement of the beam at an $x \in [0, l]$ and a $t \geq 0$ is denoted by $y(x, t)$ .

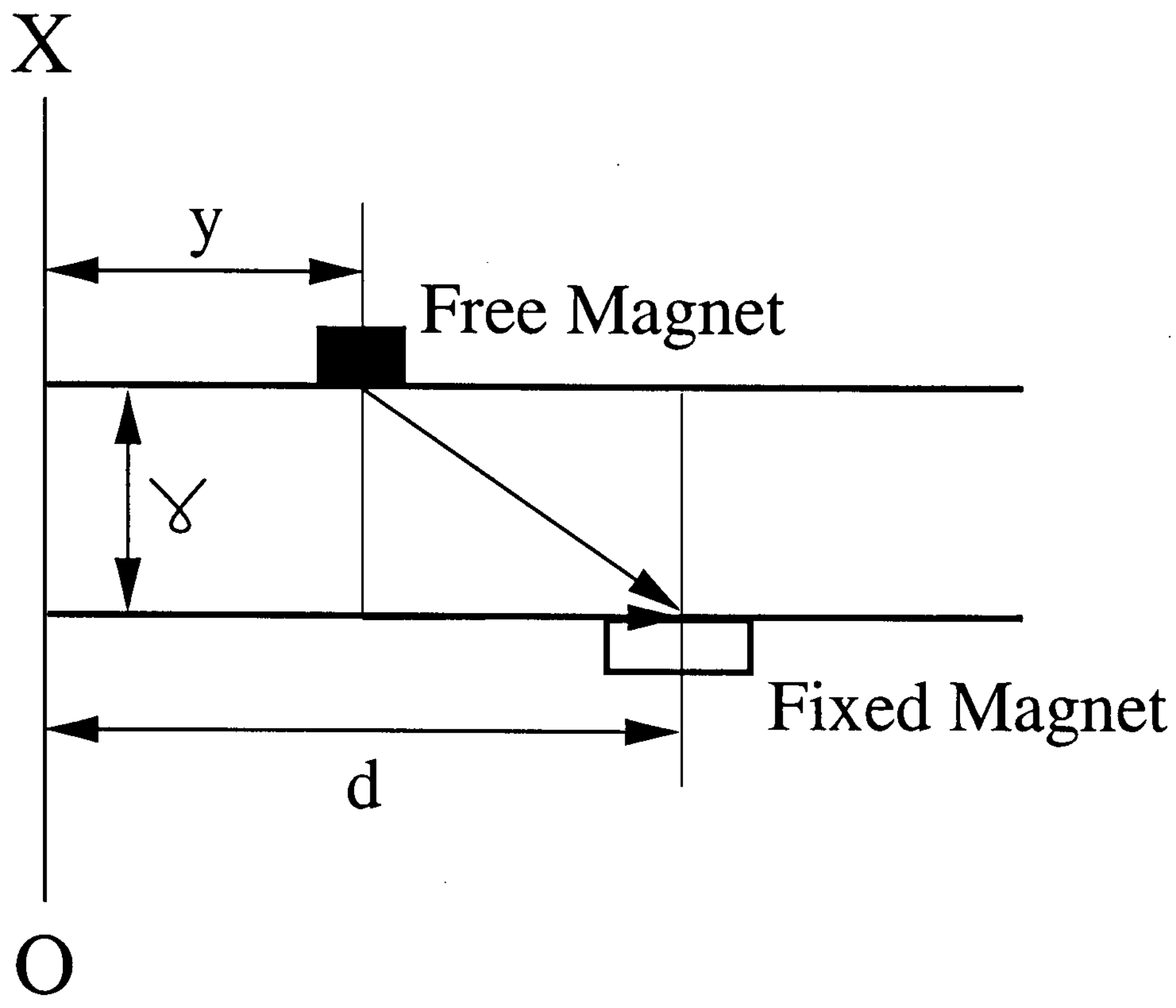

**Fig. 3:** A free magnet separated from a fixed attracting magnet by a glass plate of thickness $\gamma$ . The free magnet moves towards the fixed one.

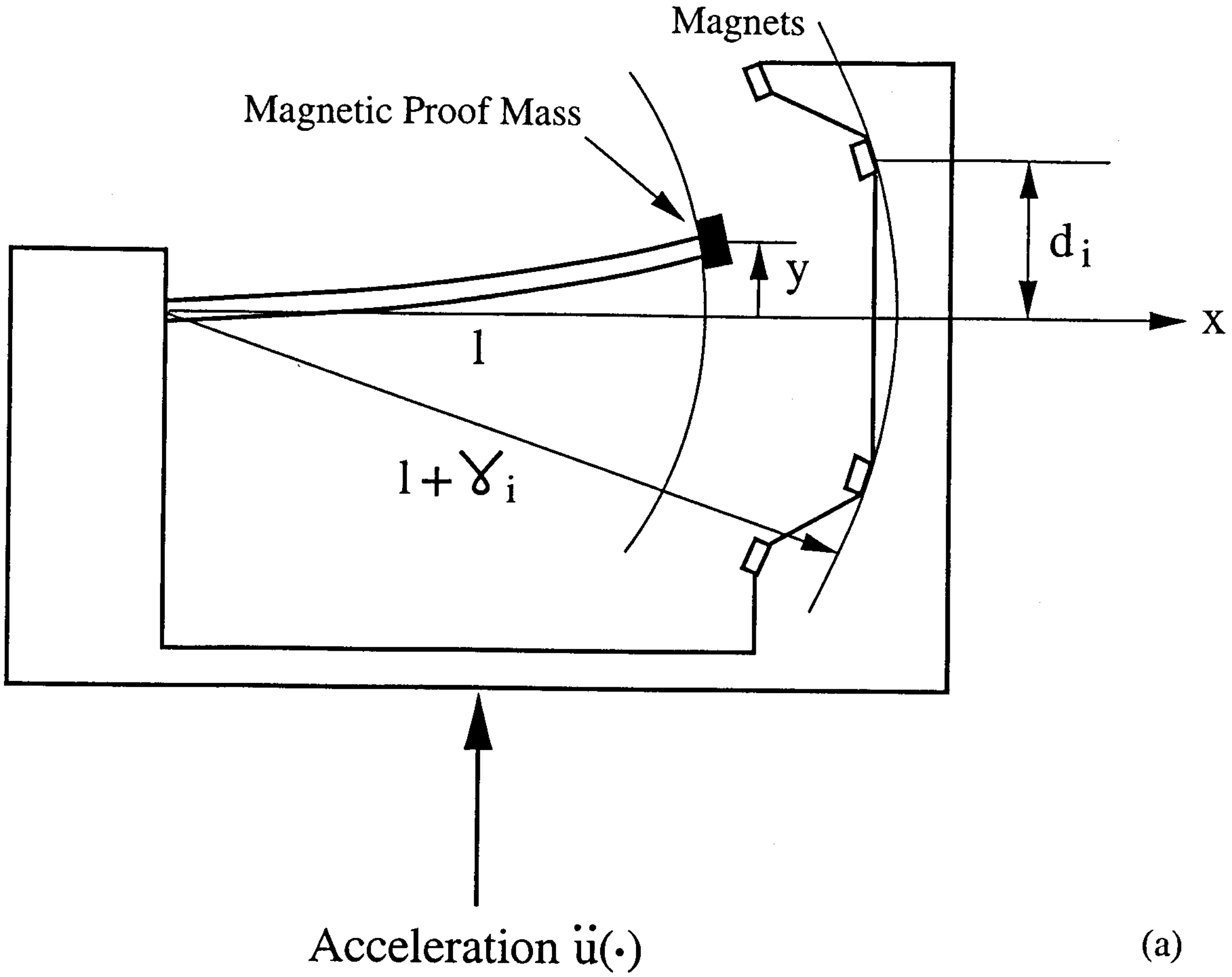

(a)

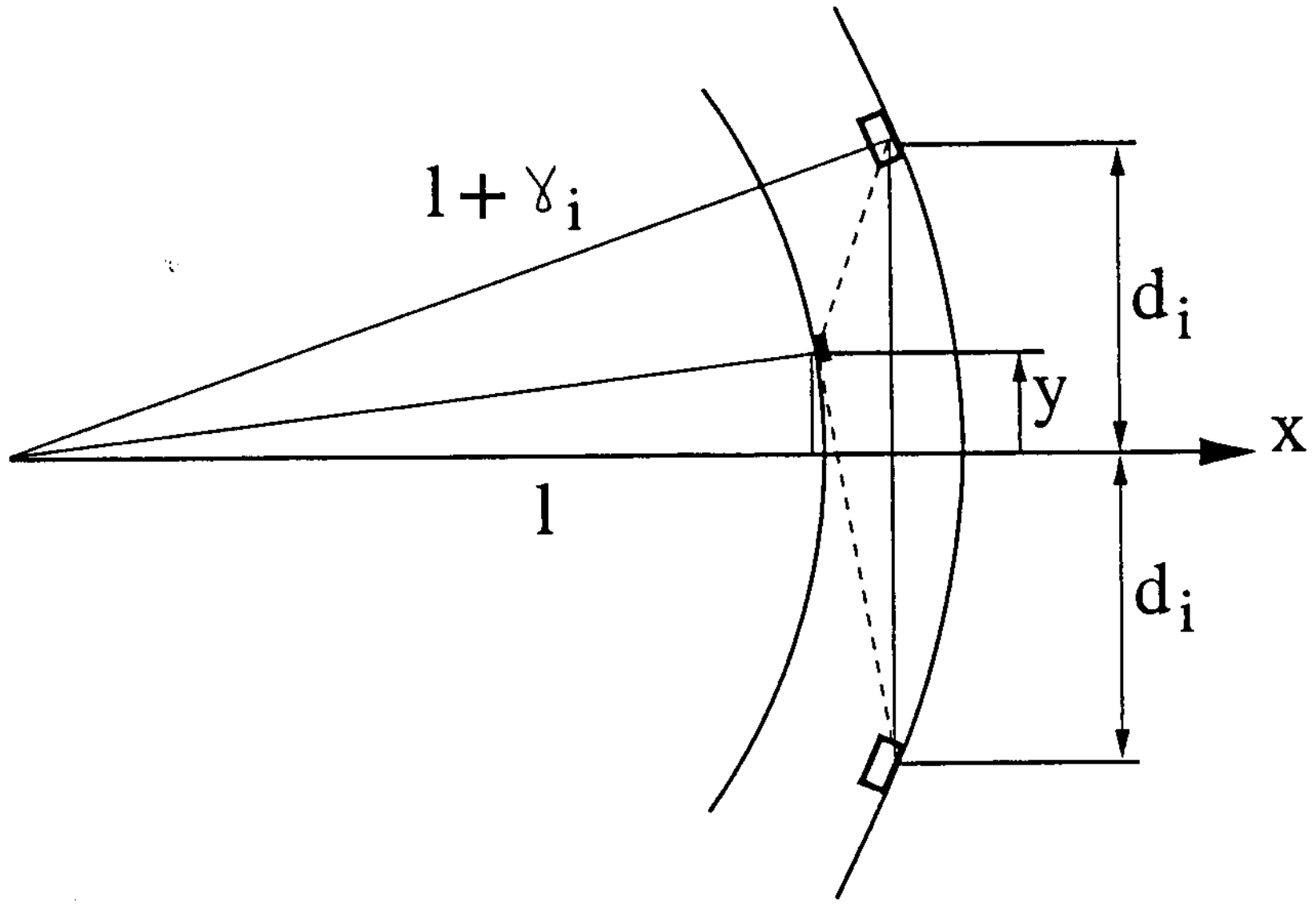

(b)

**Fig. 4:** (a) A cantilever beam with a magnetic proof mass. A collection of attracting magnets are placed in the vicinity of the beam-mass system. The magnet above and below the $x$-axis are the mirror image of each other geometrically and with respect to the strengths of the magnets; (b) The distances between the magnetic proof mass and the fixed magnets (lengths of the dashed lines) are determined using this figure.

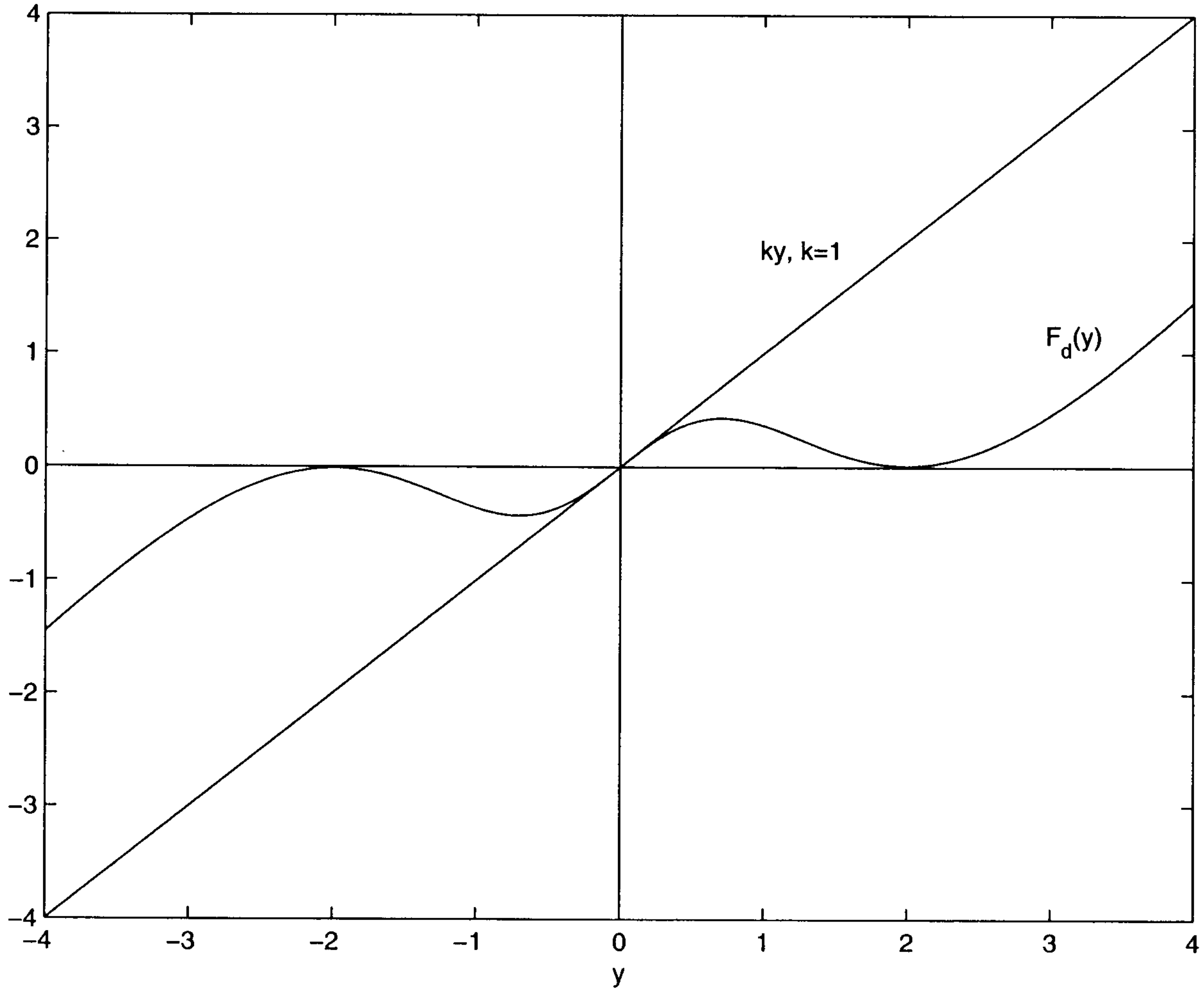

**Fig. 5:** The graph of the nonlinear spring restoring force $y \mapsto F_d(y)$ in Eq. (16) in Example 3.1 and the graph of the linear spring force $y \mapsto ky$ with $k = 1$ .

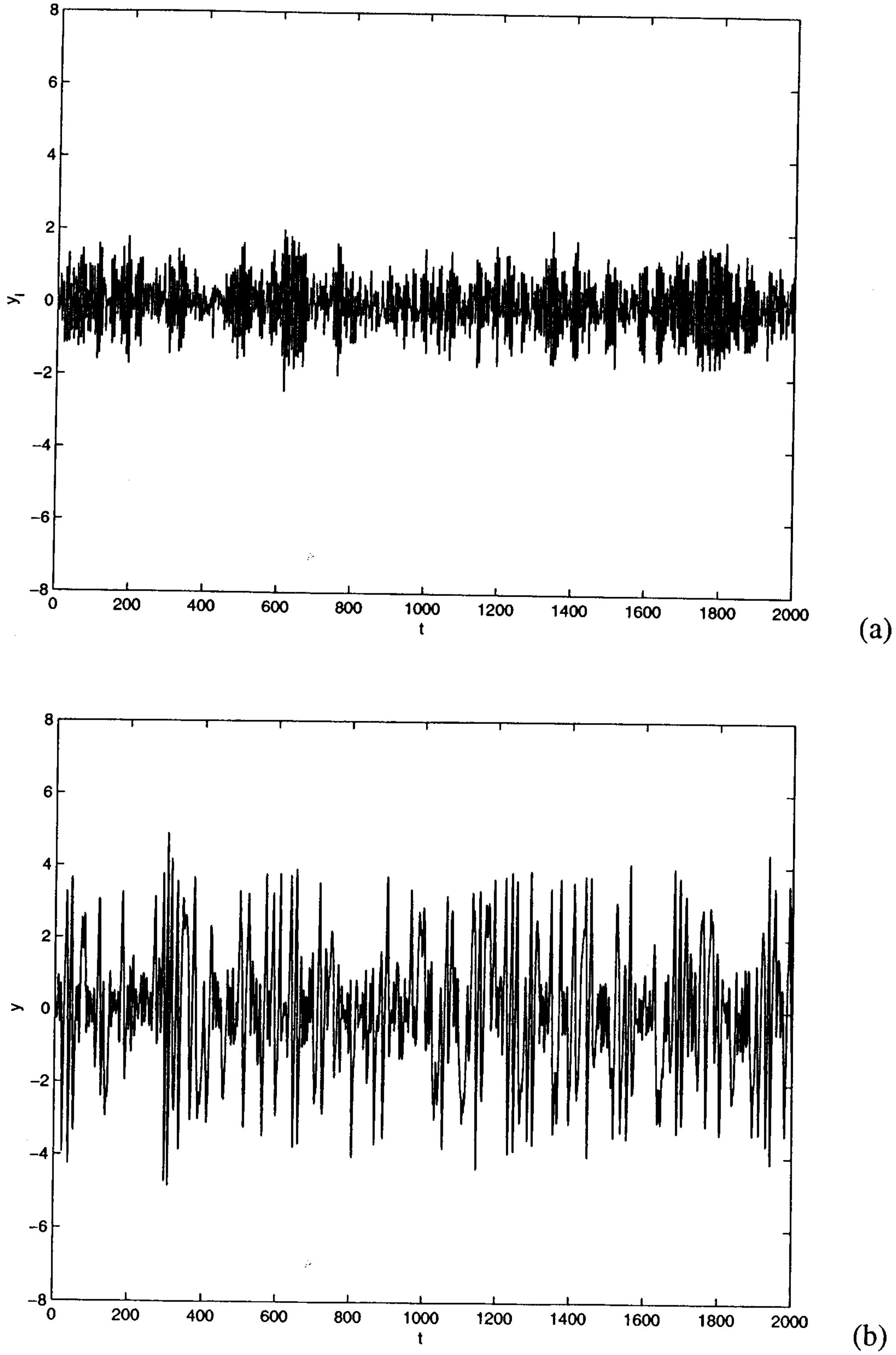

**Fig. 6:** Response of the linear beam-mass system (7), $y_l$, and that of the nonlinear beam-mass system (13), $y$, in Example 3.1, respectively, in (a) and (b). The amplitude of $y$ is four times larger than that of $y_l$.

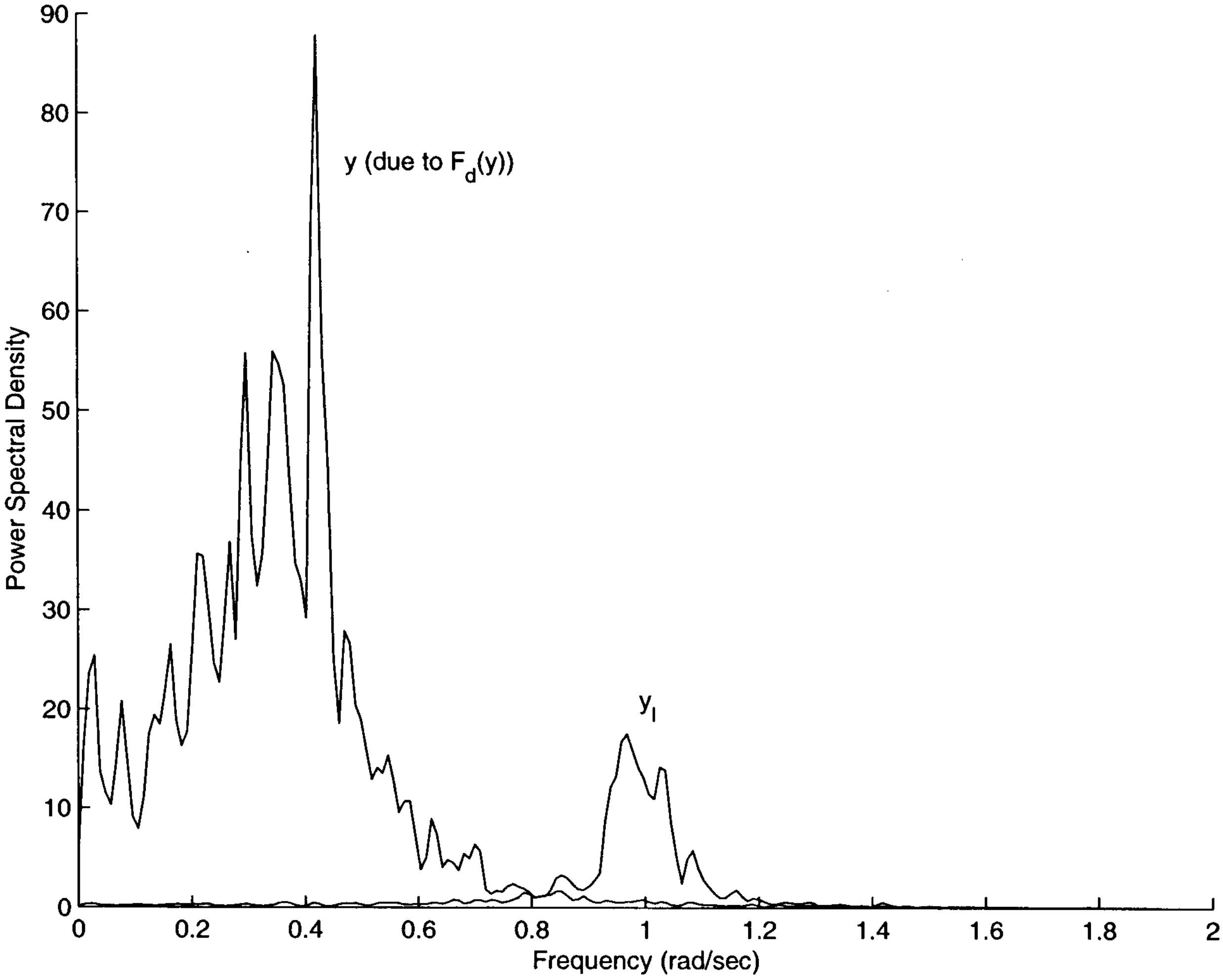

**Fig. 7:** Power spectral densities of $y_l$ and $y$ in Example 3.1.

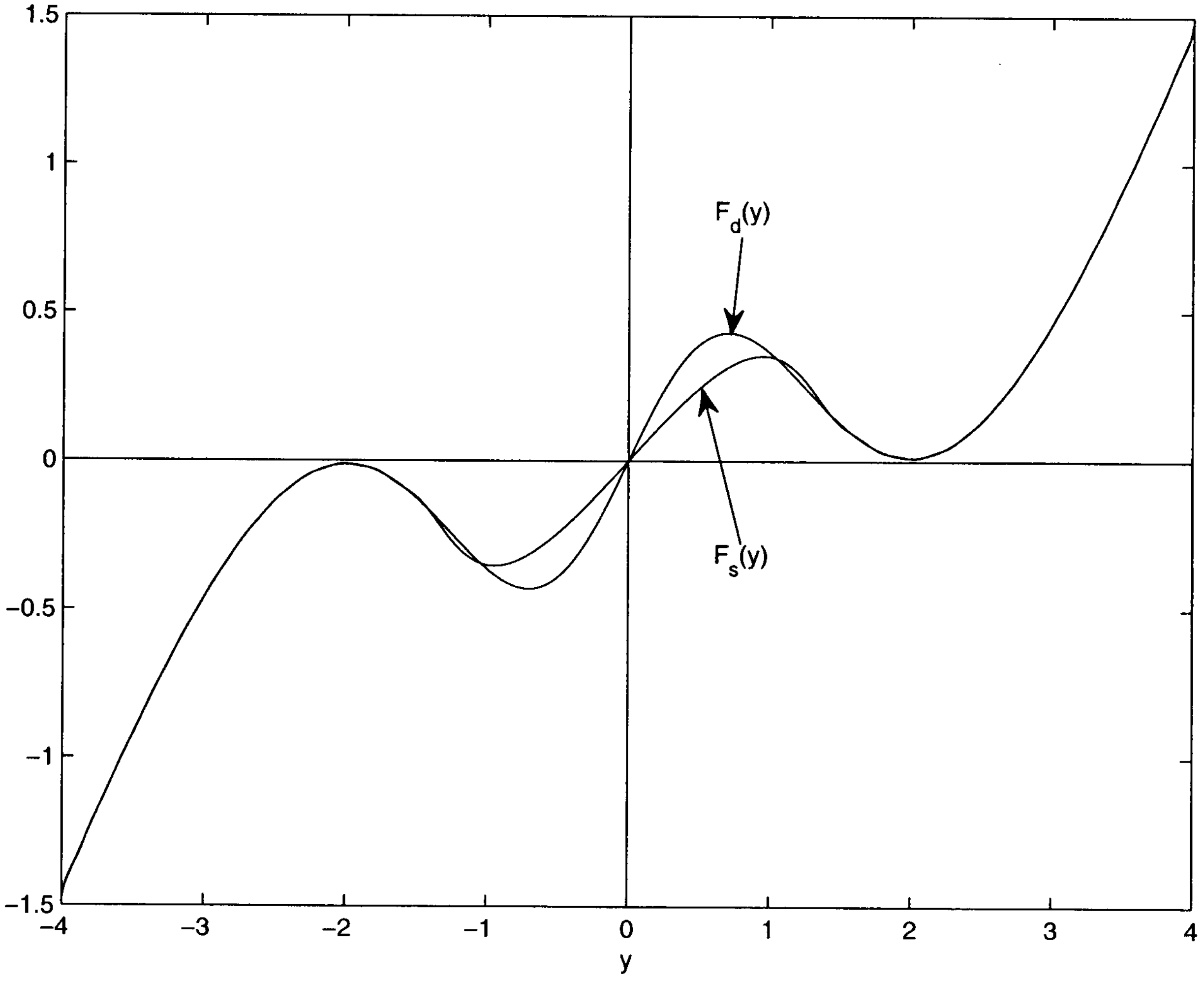

**Fig. 8:** The graph of $y \mapsto F_s(y)$ in Eq. (14) constructed in Example 3.2 compared to that of $y \mapsto F_d(y)$ in Eq. (16).

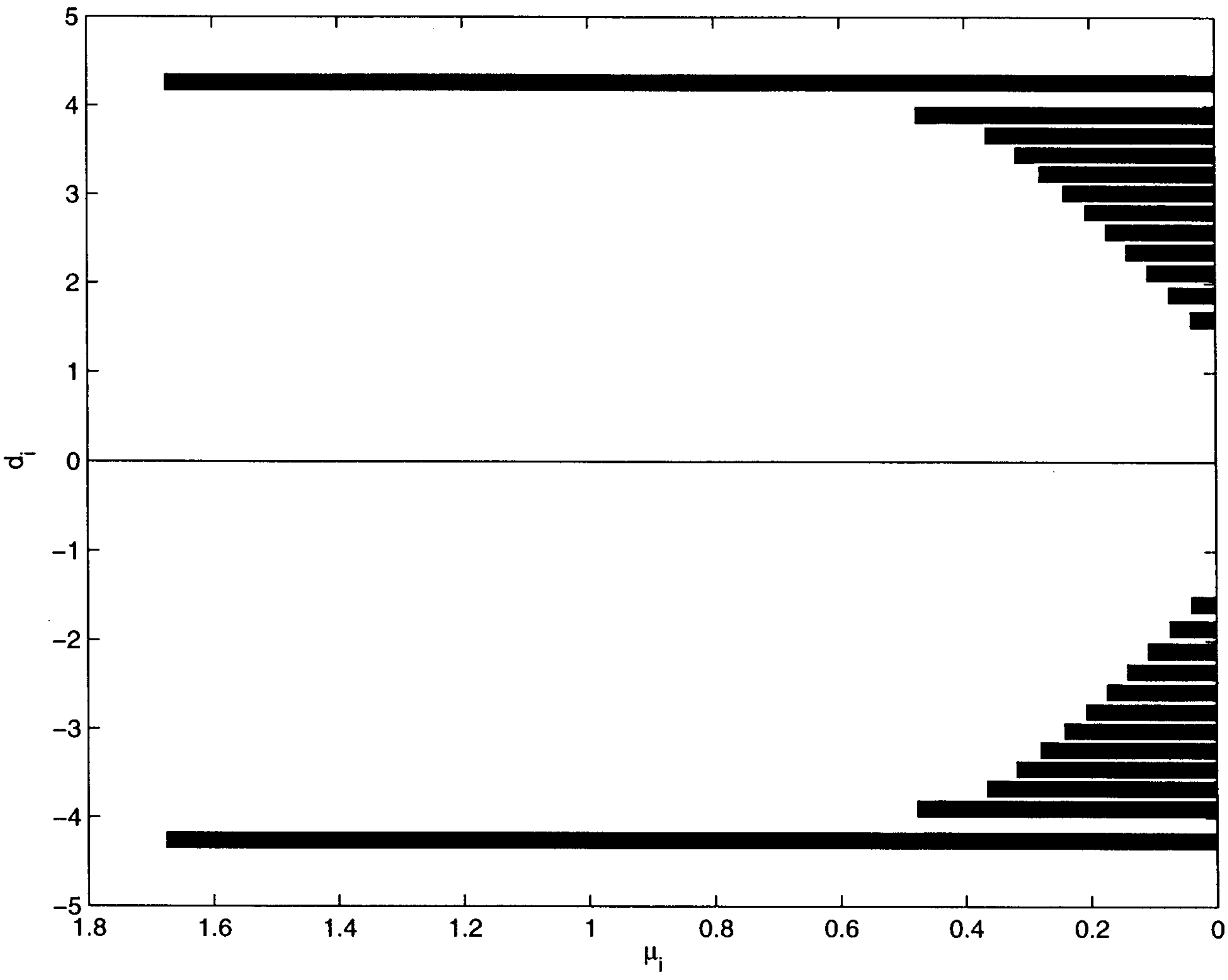

**Fig. 9:** The strengths and positions of the 24 magnets determined in Example 3.2.

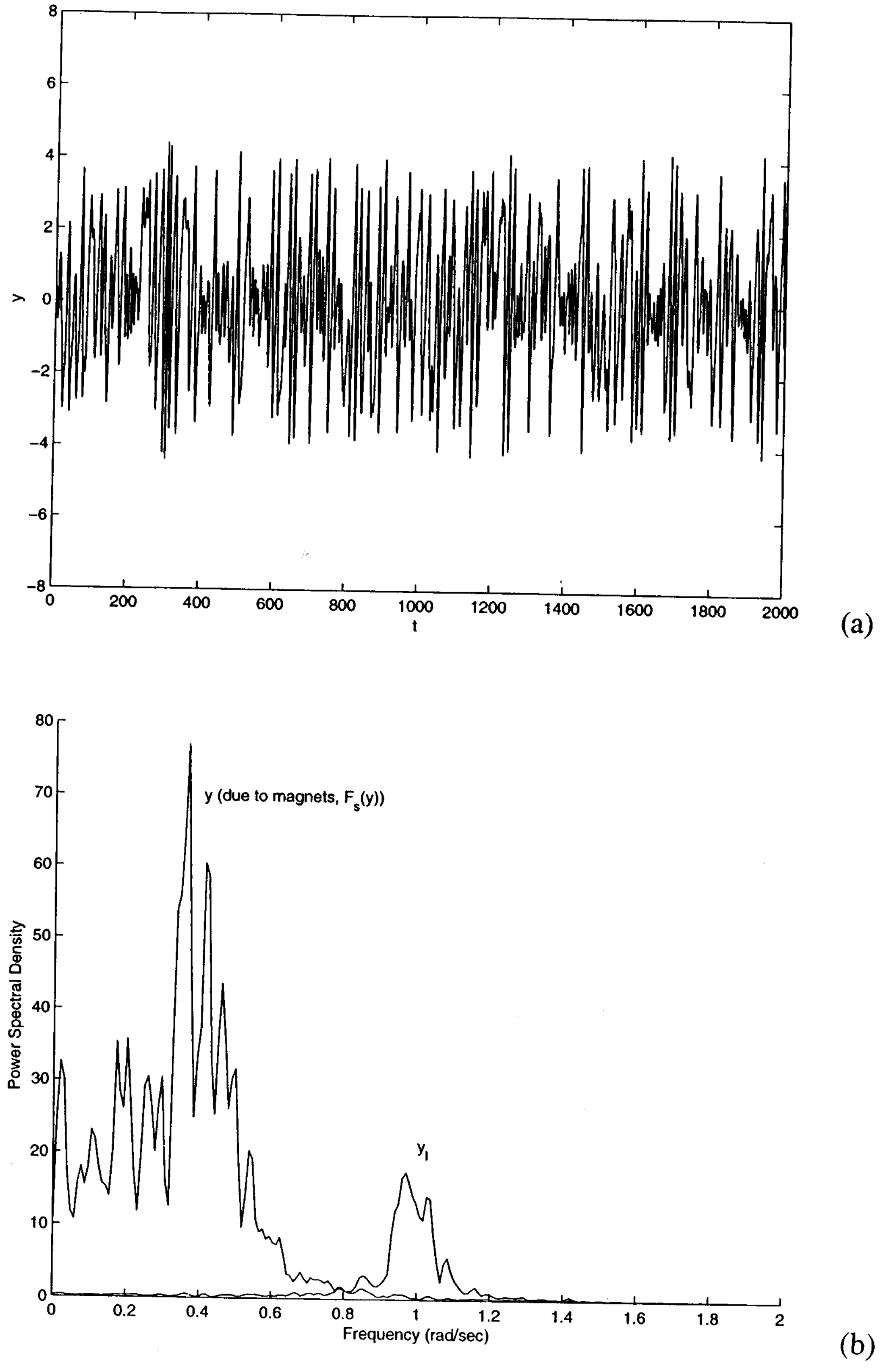

**Fig. 10:** (a) Response of system (13), $y$, in which the function $y \mapsto F_s(y)$ is that determined in Example 3.2; (b) Power spectral densities of $y_l$ and $y$.

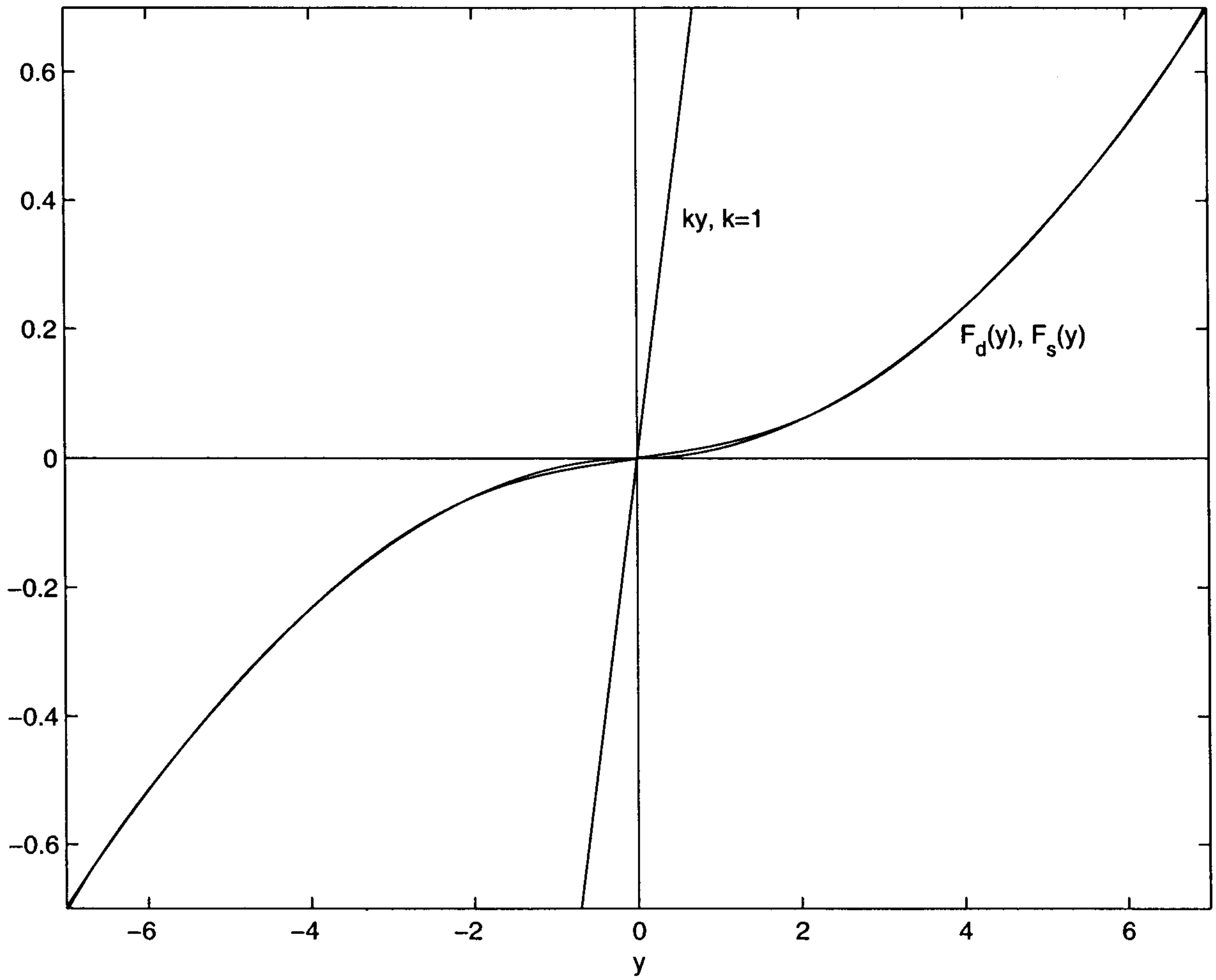

**Fig. 11:** The graphs of $y \mapsto F_d(y)$ and $y \mapsto F_s(y)$ in Example 3.4 and that of $y \mapsto ky$ with $k = 1$ .

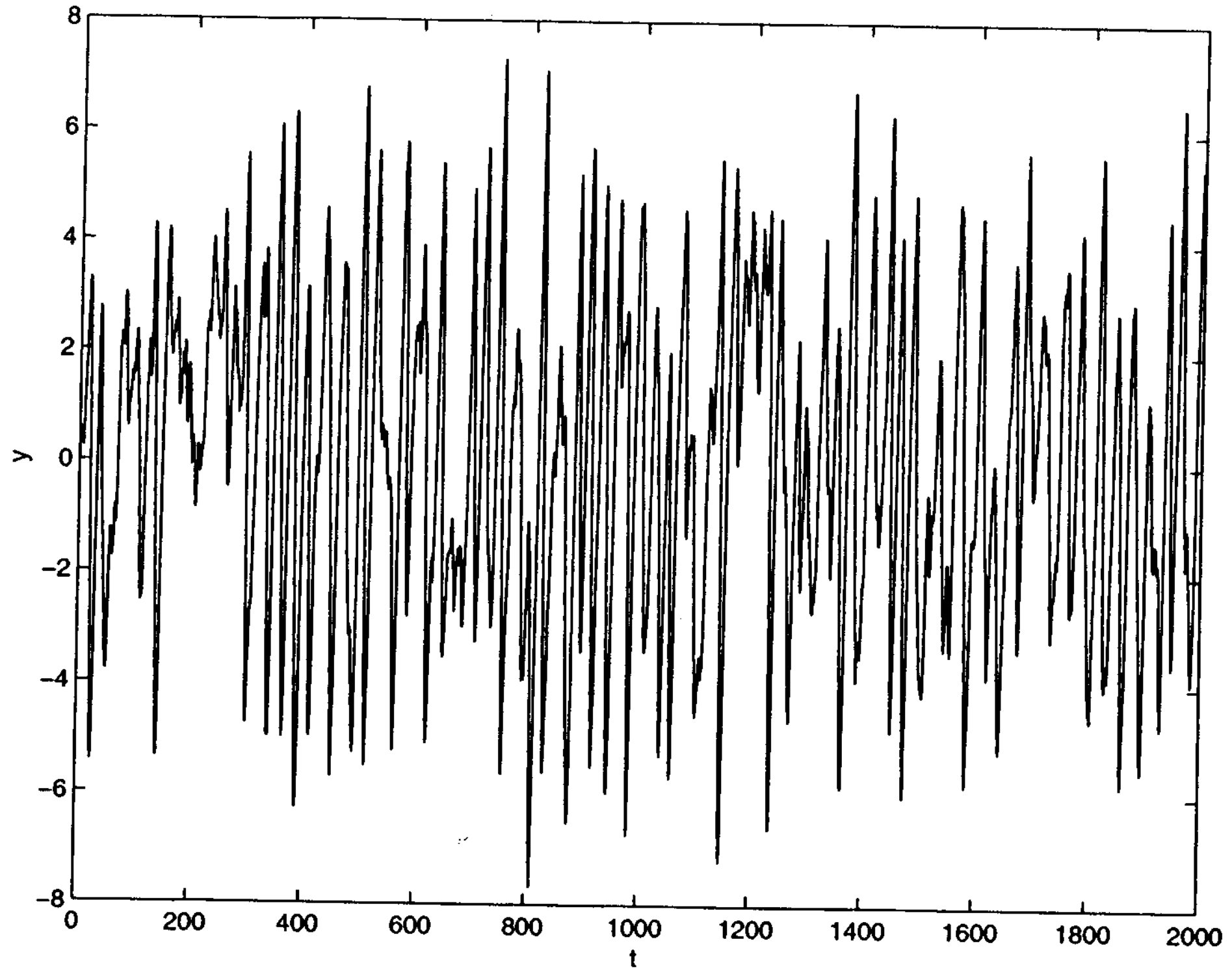

(a)

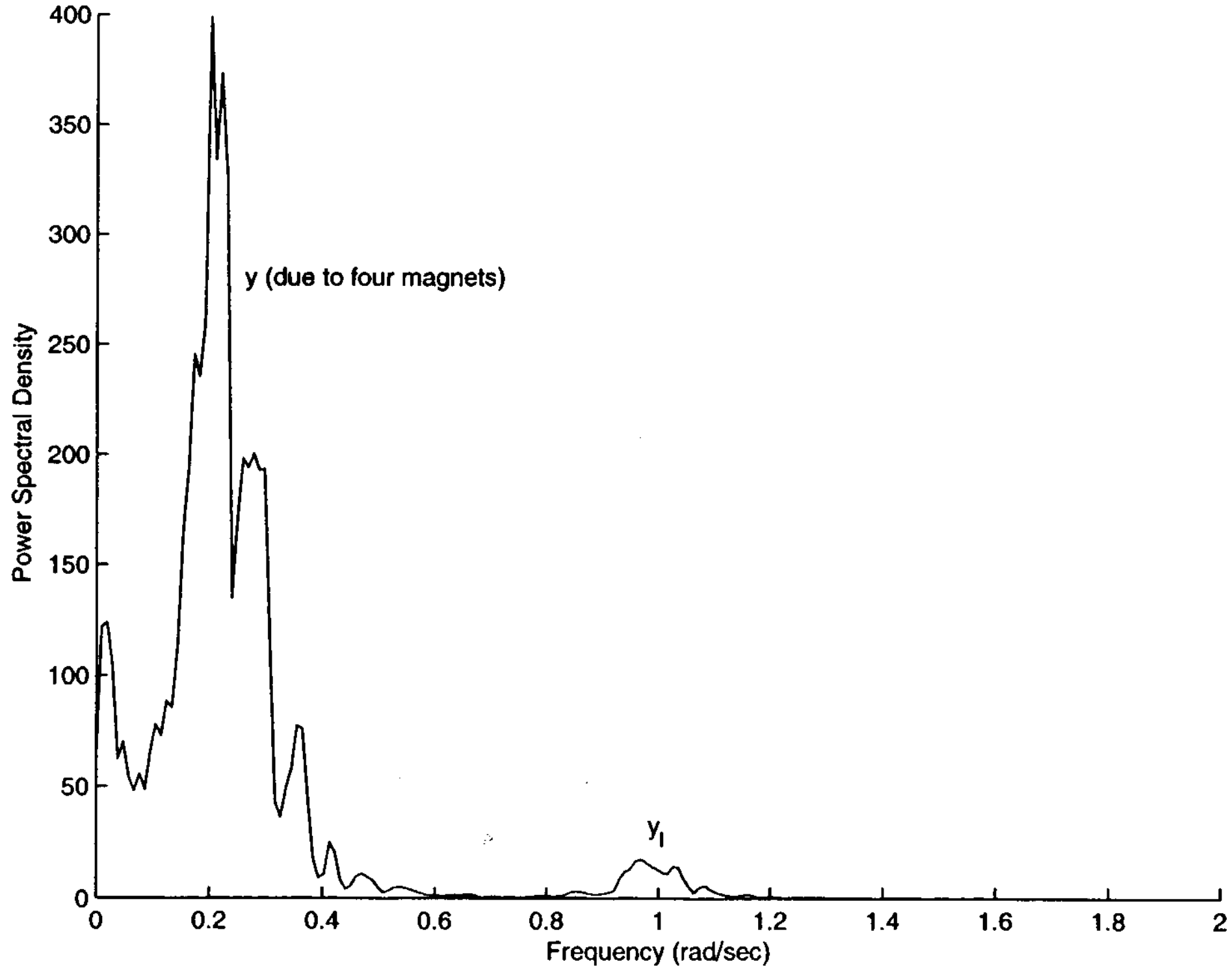

(b)

**Fig. 12:** (a) Response of system (13), $y$, in Example 3.5, where there are four magnets in the vicinity of the magnetic proof mass; (b) Power spectral densities of $y_l$ and $y$.

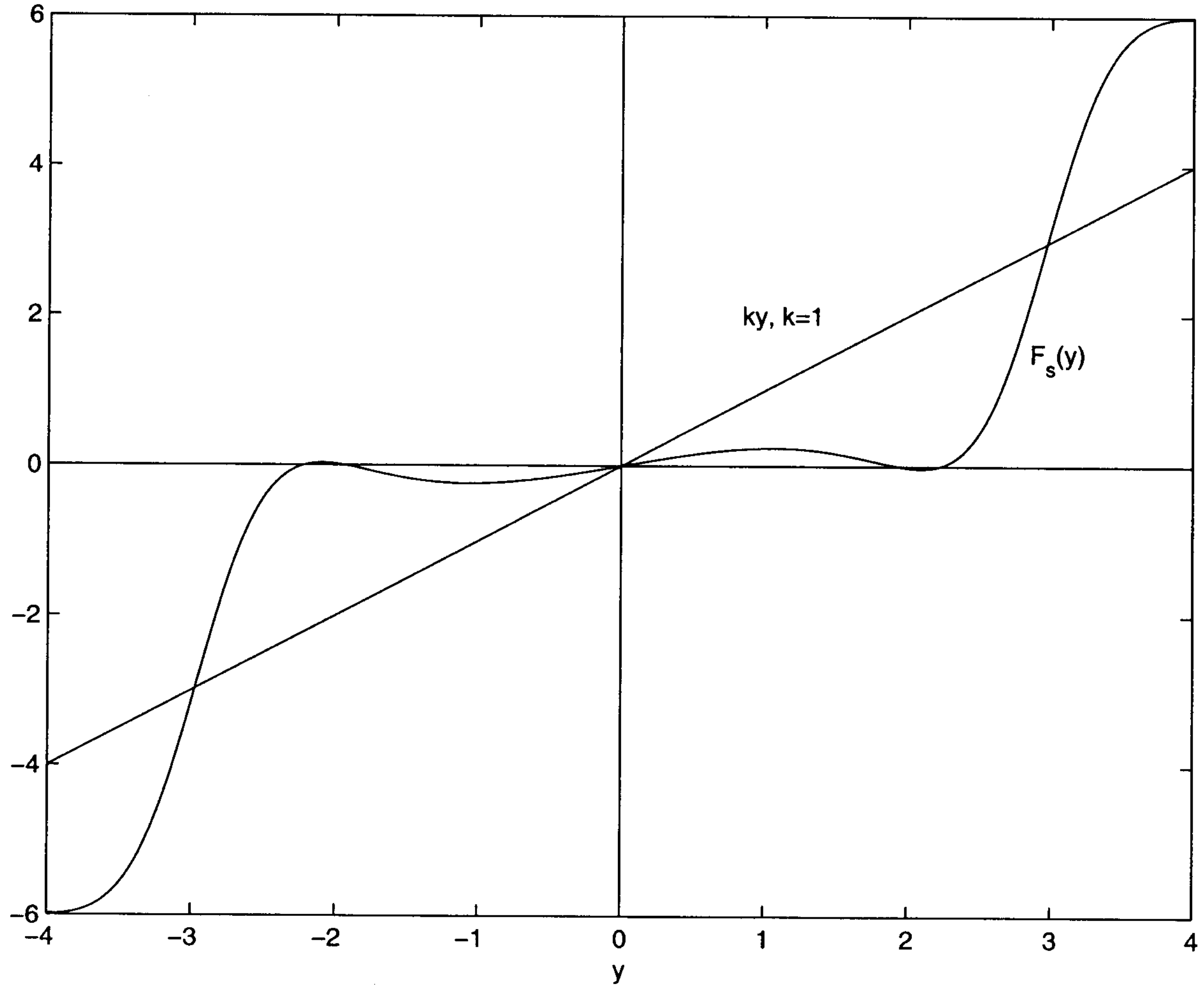

**Fig. 13:** The graph of $y \mapsto F_s(y)$ in Example 3.6 and that of $y \mapsto ky$ with $k = 1$ .

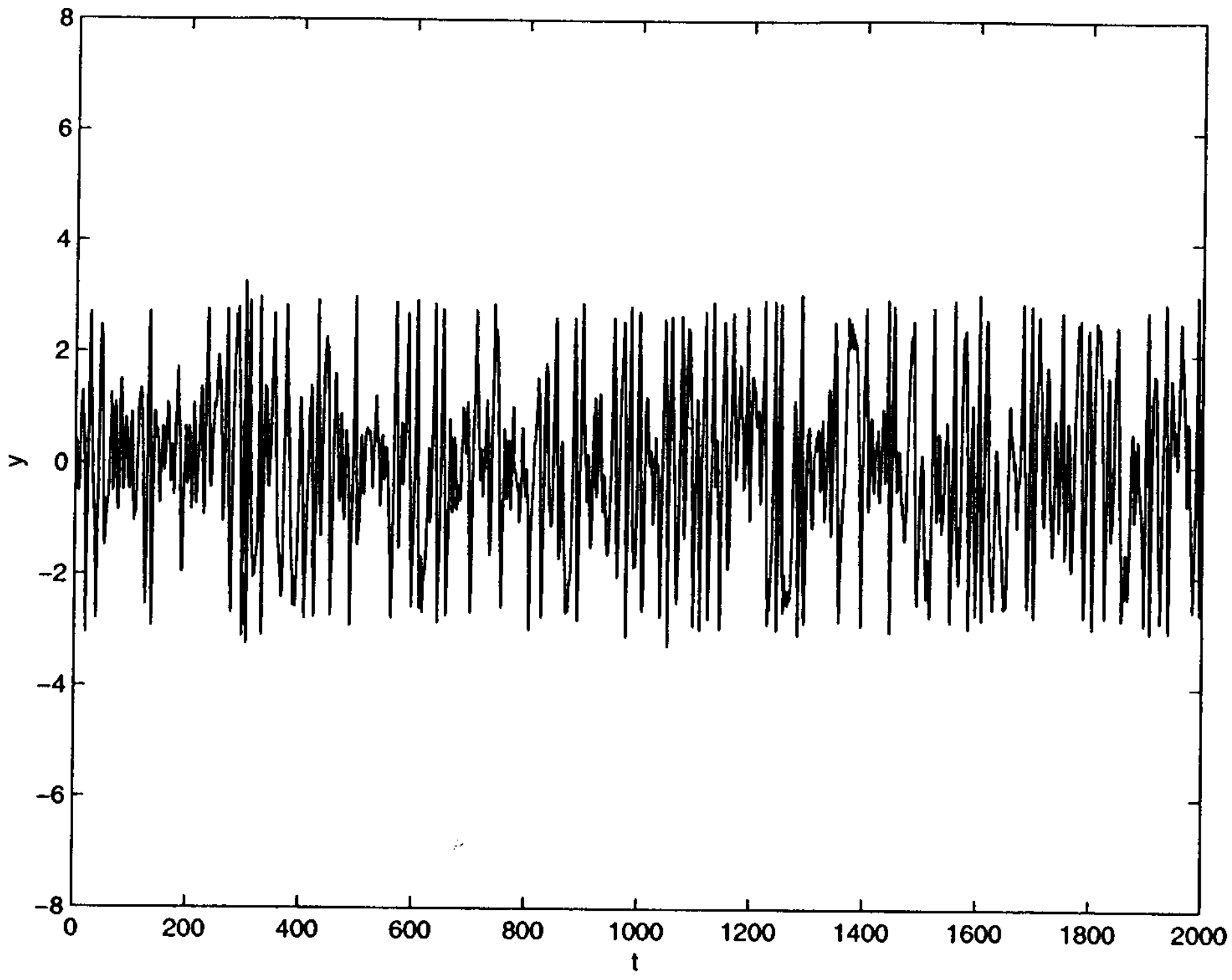

(a)

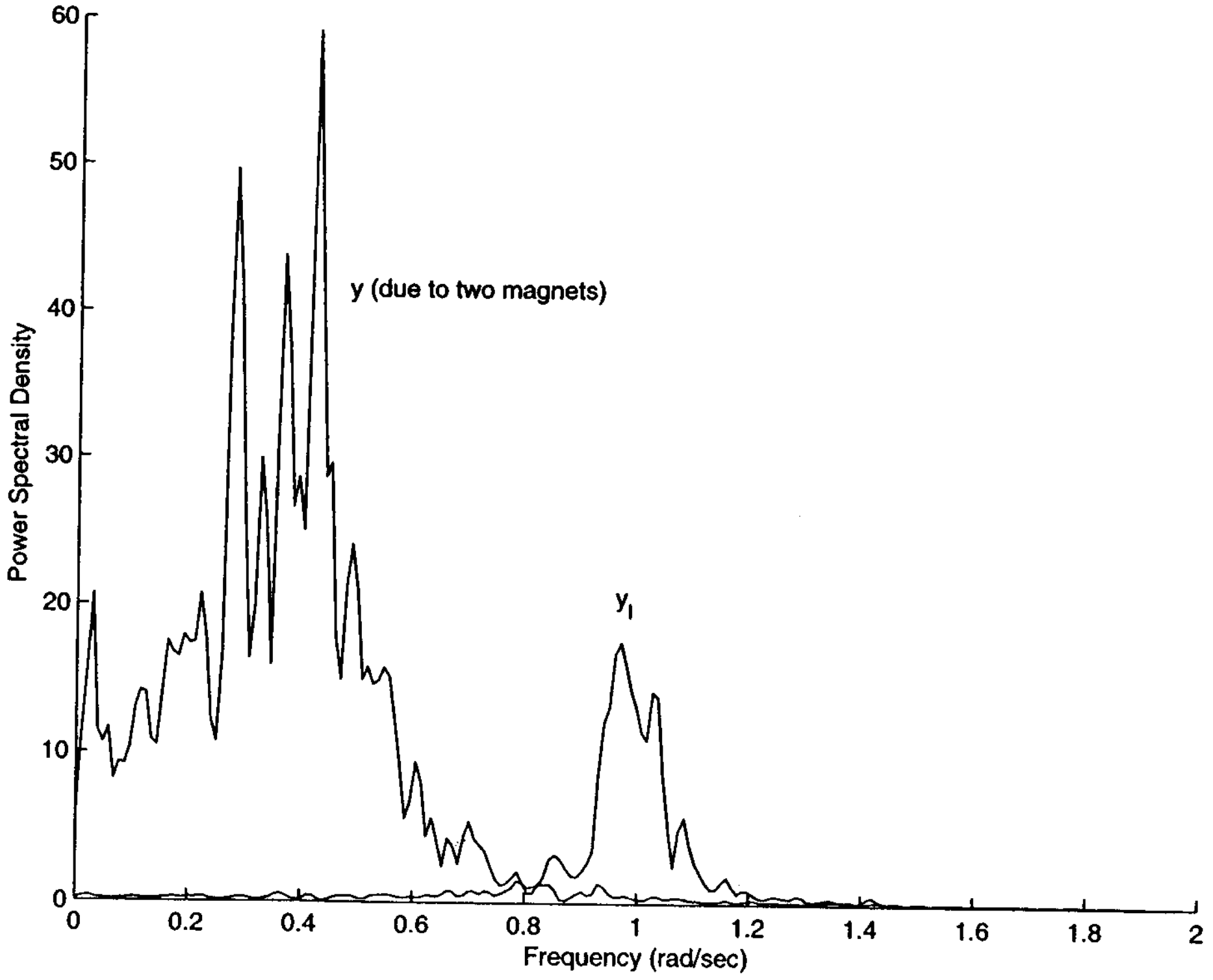

(b)

**Fig. 14:** (a) Response of system (13), $y$, in Example 3.7, where there are two magnets in the vicinity of the magnetic proof mass; (b) Power spectral densities of $y_l$ and $y$.